\documentclass[a4paper, 11pt, notitlepage]{article}
\usepackage[utf8]{inputenc}
\usepackage{epsfig}
\usepackage{float}

\usepackage[usenames,dvipsnames]{color}

\usepackage{hyperref}
\hypersetup{dvips,colorlinks,citecolor=OliveGreen,linkcolor=BrickRed}

\begin{document}

\title{Physics Beyond Standard Model\\ in Neutron Beta Decay
    \footnote{Presented by M.~Ochman at the XXXV International Conference of Theoretical Physics "Matter to the Deepest 2011", Ustro\'n, Poland, September 12--18, 2011.}
}
\author{Jacek~Holeczek, Micha\l{}~Ochman, El\.zbieta~Stephan, Marek~Zra\l{}ek\\
	\small{Institute of Physics, University of Silesia, Poland}
}
\date{}

\maketitle
\begin{abstract}
Limits from neutron beta decay on parameters describing physics beyond the Standard Model are presented. New Physics is described by the most general Lorentz invariant effective Hamiltonian involving vector, scalar and tensor operators and Standard Model fields only. Two-parameter fits to the decay parameters measured in free neutron beta decay have been done, in some cases indicating rather big dependence of the results on $g_A/g_V$ ratio of nucleon form factors at zero four-momentum transfer.
\end{abstract}
 
\section{Introduction}

For many years nuclear $\beta$-decays have been exploited as laboratories for testing the Standard Model (SM) in the domain of low energies. Along with developments of intense sources of cold and ultracold neutrons and  improvements of experimental techniques, the precision of measurements in the simplest of such systems: the $\beta$-decay of a free neutron, is constantly increasing. It opens the way to study the limits on physics  beyond SM set solely by the parameters of the neutron $\beta$-decay.

We assume that at the quark--lepton level $\beta$-decay is described by the general 4-point Hamiltonian \cite{Herczeg}
\begin{eqnarray} \label{hamiltonian}
\mathcal{H}_{\beta}\; =\; 4\sum_{k,l\, =\, L,R}\!\!
&\!\! \biggl\{ \!\! & 
\! a_{kl}\ \bar{e} \gamma_\mu P_k \nu^{(k)}\,  \bar{u} \gamma^\mu P_l d  \nonumber \\
&+&
\! A_{kl}\ \bar{e} P_k \nu^{(k)}\,  \bar{u} P_l d  \nonumber \\
&+&
\! \alpha_{kk}\ \bar{e} \frac{\sigma_{\mu\nu}}{\sqrt{2}} P_k \nu^{(k)}\,  
\bar{u} \frac{\sigma^{\mu\nu}}{\sqrt{2}} P_k d 
\ \biggl\}\ +\ \textnormal{H.c.,} 
\end{eqnarray}
where $u$, $d$ are quark fields, $e$ stands for electron field and 
$P_L = \frac{1}{2}\left( 1 - \gamma_5 \right)$, $P_R = \frac{1}{2}\left( 1 + \gamma_5 \right)$,
$\sigma_{\mu\nu} = \frac{i}{2}\left[\gamma_\mu,\gamma_\nu\right]$, where our metric and gamma matrices are the same as {\it e.g.} in \cite{Giunti}. We work in the basis in which mass matrix of charged leptons is diagonal and the left~($L$) and right~($R$) neutrino fields are given by
\begin{eqnarray}
\nu^{(L)} & = & \sum_i \mathrm{U}_{ei} P_L \nu_i\, \textnormal{,} \\
\nu^{(R)} & = & \sum_i \mathrm{V}_{ei} P_R \nu_i\, \textnormal{,}
\end{eqnarray}
where $\nu_i$ is the $i$-th neutrino field with a certain mass, $\mathrm{U}$ and $\mathrm{V}$ are respectively  the Pontecorvo--Maki--Nakagawa--Sakata matrix and similar mixing matrix for right-handed neutrinos.  SM is restored when $a_{kl}=A_{kl}=\alpha_{kk}=0$ for $k,\,l = L,\,R$ except $a_{LL} = V_{ud}\, G_F/\sqrt{2}$, where $G_F$ is the usual Fermi constant and $V_{ud}$ is the element of quark Cabibbo--Kobayashi--Maskawa mixing matrix.

When calculating the amplitudes for neutron beta decay at small four-momentum transfer $q^2 \approx 0$ we have used the relations \cite{Herczeg}
\begin{eqnarray}
g_{V}\bar{u}_{p} \gamma_\mu u_{n} & = & \langle p | \bar{u}
\gamma_\mu d | n \rangle\textnormal{,}  \\
g_{A}\bar{u}_{p} \gamma_\mu \gamma_5 u_{n} & = & \langle p |
\bar{u} \gamma_\mu \gamma_5 d | n \rangle\textnormal{,}  \\
g_{S}\bar{u}_{p} u_{n} & = & \langle p | \bar{u} d | n \rangle\textnormal{,}  \\
g_{T}\bar{u}_{p} \sigma_{\mu\nu} u_{n} & = & 
\langle p | \bar{u} \sigma_{\mu\nu} d | n \rangle\textnormal{,} 
\end{eqnarray}
with $\langle p |$, $u_{p}$ and $| n \rangle$, $u_{n}$ being proton and neutron states and bispinors, respectively. From conserved vector current hypothesis one gets $g_{V} = 1$. In the quark model with spherically symmetric wave functions of quarks the following relations have been derived \cite{wsp_g}: $g_{S} = - \frac{1}{2} + \frac{9}{10} g_{A}$ and $g_{T} = \frac{5}{3} \left( \frac{1}{2} + \frac{3}{10} g_{A} \right)$. Substituting the SM value for $g_{A} \simeq 1.27$ into the above relations leads to: $g_{S} \simeq 0.64$ and $g_{T} \simeq  1.47$. However, in our derivations and fits we treat $g_{S}$ and $g_{T}$ as free parameters (independent of $g_{A}$).

\section{Decay Parameters}

From Eq.~(\ref{hamiltonian}) the five-fold differential decay width for polarized neutron without measurement of final electron and proton polarization is given by (in analogy to \cite{Jackson})
\begin{eqnarray} \label{d_Gamma}
\frac{d\Gamma}{dE_e d\Omega_e d\Omega _\nu}
\ \sim \ p_e E_e E_\nu^2\,
\!\! & \!\! \bigg\{ \!\! & \!\! 
1 + a \frac{\vec{p}_e \cdot \vec{p}_\nu}{E_e E_\nu} + 
b \frac{m_e}{E_e} \nonumber \\
& & 
\!\! +\, \vec{\lambda}_n \cdot \left[ A
\frac{\vec{p}_e}{E_e} + B \frac{\vec{p}_\nu}{E_\nu} + D
\frac{\vec{p}_e \times \vec{p}_\nu}{E_e E_\nu} \right]
\bigg\}\,\textnormal{,}
\end{eqnarray}
where $\vec{\lambda}_n$ is the neutron polarization vector, $m_e$, $p_e = |\vec{p}_e|$, $E_e$ are, respectively, the mass, momentum and total energy of electron, $E_0$ is the maximum value of $E_e$, $|\vec{p}_\nu| = E_\nu = E_0 - E_e$ is the antineutrino energy\footnote{The effect of nonzero neutrino masses enters only trough presence of mixing matrices $U$ and $V$.}. The $\Omega_e$, $\Omega _\nu$ denotes the solid angles of electron and antineutrino emission. We have worked at tree-level (except: calculation of $\langle E_e^{-1} \rangle$ --- see below) and with approximations such that terms proportional to $\bar{u}_{p} \gamma_5 u_{n}$ are not present in~(\ref{d_Gamma}). Furthermore, we will consider only cases when: $g_V$, $g_A$, $g_S$, $g_T$, as well as $a_{kl}$, $A_{kl}$, $\alpha_{kk}$ for $k,\,l = L,\,R$ are real --- then $D \equiv 0$ and time reversal symmetry is preserved, that is well motivated experimentally (PDG average \cite{PDG}: $D=(-4 \pm 6) \times 10^{-4}$). 

We express the decay parameters $a$, $b$, $A$, $B$, where $B$ has the form of $B = B_{0} + b_\nu\, m_e / E_e$, in terms of the ratio $g_A/g_V$ and the following parameters (see also \cite{Herczeg}) for $k,\,l = L,\,R$
\begin{equation} \label{VST}
V_{kl} \ = \ \frac{ a_{kl} }{ a_{LL} } \kappa_k \,\textnormal{,} \qquad
S_{kl} \ = \ \frac{ A_{kl} }{ a_{LL} } \frac{g_S}{g_V} \kappa_k \,\textnormal{,} \qquad
T_{kl} \ = \ \frac{ \alpha_{kl} }{ a_{LL} }\frac{g_T}{g_V} \kappa_k \,\textnormal{,}
\end{equation}
where
\begin{equation} \label{kappa_LR}
\kappa_L\ =\ 1\,\textnormal{,} \qquad 
\kappa_R\ =\ \left( \frac{\sum_i' | \mathrm{V}_{ei} |^2}
                               {\sum_i' | \mathrm{U}_{ei} |^2} \right)^{1/2}\,\textnormal{,}
\end{equation}
with summation $\sum_i'$ running only over kinematically allowed antineutrino states. In SM and for some cases of physics beyond SM $b = 0$ and $b_\nu = 0$. As a result of the applied approximations formulas for $a$, $b$, $A$, $B_{0}$, $b_\nu$ depend in general case only on two combinations 
\begin{eqnarray}
s_L & = & S_{LL} + S_{LR}\,\textnormal{,}\\
s_R & = & S_{RR} + S_{RL}\,\textnormal{.} 
\end{eqnarray}
Next, following the approach applied in \cite{Gluck,Severijns,Konrad}, we define
\begin{eqnarray}
\bar{a}( \langle W^{-1}\rangle ) & = & \frac{ a }{ 1 + b \langle W^{-1}\rangle }\, \textnormal{,} \\	
\bar{A}( \langle W^{-1}\rangle ) & = & \frac{ A }{ 1 + b \langle W^{-1}\rangle }\, \textnormal{,} \\
\bar{B}( \langle W^{-1}\rangle ) & = & \frac{ B_{0} + b_\nu \langle W^{-1}\rangle }
                                                    { 1 + b \langle W^{-1}\rangle }\, 
\textnormal{,} 
\end{eqnarray}
where $\langle W^{-1}\rangle = m_e\langle E_e^{-1} \rangle$, and apply these quantities in the fits to the experimental data. The $\chi^2$, which we will minimize with the fit procedure, is of the form
\begin{eqnarray} \label{chi2}
\chi^2 	& = &   \sum_i \left[\frac{a_i - \bar{a}(\langle W^{-1}\rangle_i)}{\delta a_i}\right]^2
				\nonumber \\
		& + &   \sum_j \left[\frac{A_j - \bar{A}(\langle W^{-1}\rangle_j)}{\delta A_j}\right]^2 
				\nonumber \\
		& + &   \sum_k \left[\frac{B_k - \bar{B}(\langle W^{-1}\rangle_k)}{\delta B_k}\right]^2 
				\,\textnormal{,}
\end{eqnarray}
where the selected data are presented in Table~\ref{data_table}: $a_i$, $A_j$, $B_k$ and $\delta a_i$, $\delta A_j$, $\delta B_k$ denote the central value and the error of the respective decay parameter in a certain experiment. We calculate the particular value of $\langle W^{-1}\rangle_i = m_e\langle E_e^{-1} \rangle_i$ from
\begin{equation}
\langle E_e^{-1} \rangle_i \ =\  
{\displaystyle \int_{E_i^{min}}^{E_i^{max}} d E_e \frac{d\Gamma}{dE_e} E_e^{-1}} \bigg/
       {\displaystyle \int_{E_i^{min}}^{E_i^{max}} d E_e \frac{d\Gamma}{dE_e}}
\,  \textnormal{,}
\end{equation}
where $E_i^{min}$ and $E_i^{max}$ in general are different for different experiments. At  this stage of calculation Fermi function $F(E_e)$ (that is a leading order QED correction)  \cite{Fermi,Schopper,on_continuum} has been incorporated and SM was assumed
\begin{equation}
  \frac{d\Gamma}{dE_e}\ =\ (g_V^2 + 3 g^2_A)\, \frac{G^2_F |V_{ud}|^2}{2
    \pi^3} p_e E_e (E_0 - E_e)^2 F(E_e)  \,  \textnormal{,}
\end{equation}
\begin{equation}
F(E_e)\ =\ \frac{2\pi\alpha\, E_e / p_e}
{\displaystyle 1 - e^{\textstyle -2\pi \alpha\,
E_e/p_e}} \,  \textnormal{.}
\end{equation}
\begin{table}
\centering
\begin{tabular}{cllllll}
PAR.  &  VALUE  &  ERROR  &  $\langle W^{-1} \rangle$  &  PAPER ID (PDG) &  &   \\ \hline
  &     &     &     &     &  &   \\  
$a$  &  $-0.1054$  &  $0.0055$  &  $0.655$  &  BYRNE      & 02  &  \cite{BYRNE_02}  \\
 &  $-0.1017$  &  $0.0051$  &  $0.655$  &  STRATOWA   & 78   &  \cite{STRATOWA_78}  \\
 &       &      &      &             &      &  \\
$A$  &  $-0.11966$  &   $0.00166$  &   $0.557$  &  LIU        & 10   &  \cite{LIU_10}  \\
 &  $-0.1189$  &  $0.0007$  &  $0.534$  &  ABELE      & 02   &  \cite{ABELE_02}  \\
 &  $-0.1160$  &  $0.0015$  &  $0.582$  &  LIAUD      & 97   &  \cite{LIAUD_97}  \\
 &  $-0.1135$  &  $0.0014$  &  $0.558$  &  YEROZOLIMSKY  &    97   &  \cite{YEROZOLIMSKY_97}  \\
 &  $-0.1146$  &  $0.0019$  &  $0.581$  &  BOPP       & 86   &  \cite{BOPP_86}  \\
 &       &      &      &             &      &  \\
$B$  &  $0.980$  &  $0.005$  &  $0.599$  &  SCHUMANN   & 07   &  \cite{SCHUMANN_07}  \\
 &  $0.967$  &  $0.012$  &  $0.600$  &  KREUZ      & 05   &  \cite{KREUZ_05}  \\
 &  $0.9801$  &  $0.0046$  &  $0.594$  &  SEREBROV   & 98   &  \cite{SEREBROV_98}  \\
 &  $0.9894$  &  $0.0083$  &  $0.554$  &  KUZNETSOV  & 95   &  \cite{KUZNETSOV_95}  \\
&     &     &     &     &  &  
\end{tabular}
\caption{We have followed the PDG \cite{PDG} data selection but took only the most precise data (the error in measurements of $a$ is less than $ 6\% $ of central value, for $A$ and $B$ --- it is less than $ 2\% $). When experiment report statistic and systematic error separately we add these two errors in quadrature. In the case of asymmetric errors we have taken the larger of the reported errors. Most of presented values of the $\langle W^{-1} \rangle$ have been taken from \cite{Severijns}. We have used all 11 "data points" in the table above in every fit presented in this paper. Because of unsolved experimental ambiguity of neutron lifetime measurements (see \cite{PDG}) we have not included this quantity in our analyzes. \label{data_table}}
\end{table}

\newpage

\section{Results}

In SM the formulas derived for decay parameters depend on $\lambda = g_A/g_V$ alone and simplify to
\begin{equation} \label{SM_kAB}
a\, =\, \frac{1-\lambda^2}{3\lambda^2+1}\, \textnormal{,} \qquad
A\, =\, \frac{2\lambda(1-\lambda)}{3\lambda^2+1}\, \textnormal{,} \qquad
B\, =\, \frac{2\lambda(\lambda+1)}{3\lambda^2+1}\, \textnormal{.}
\end{equation}
In this case, the one-parameter fit is performed, which results in $\chi^2_{min} = 25.42$ with
\begin{eqnarray} 
                    & \pm & 0.0014\ (\,68.27\%\,\textnormal{C.L.}\,)  \label{g_A_fit_68}\\
\lambda\ =\ 1.2703 & \pm & 0.0023\ (\,90\%\,\textnormal{C.L.}\,)     \label{g_A_fit_90}\\
                    & \pm & 0.0028\ (\,95.45\%\,\textnormal{C.L.}\,)  \label{g_A_fit_95}
\end{eqnarray}
that is in a good agreement with the PDG average \cite{PDG}: $\lambda\, =\, 1.2701\, \pm\, 0.0025$ (error scaled by PDG by $1.9$). 

The above results (\ref{g_A_fit_68}--\ref{g_A_fit_95}) apply also when: $V_{kl}=S_{kl}=T_{kk}=0$ for $k,\,l = L,\,R$ except $V_{LR}$ (and $V_{LL} = 1$ by definition --- see Eqs. (\ref{VST}) and (\ref{kappa_LR})). In this case the formulas (\ref{SM_kAB}) hold for modified $\lambda$
\begin{equation}
\lambda\ =\ \frac{g_A}{g_V}\, \frac{1 - V_{LR}}{1 + V_{LR}} \, \textnormal{.}
\end{equation}

\begin{figure}[H]
\centering

\includegraphics[width=0.48\textwidth]{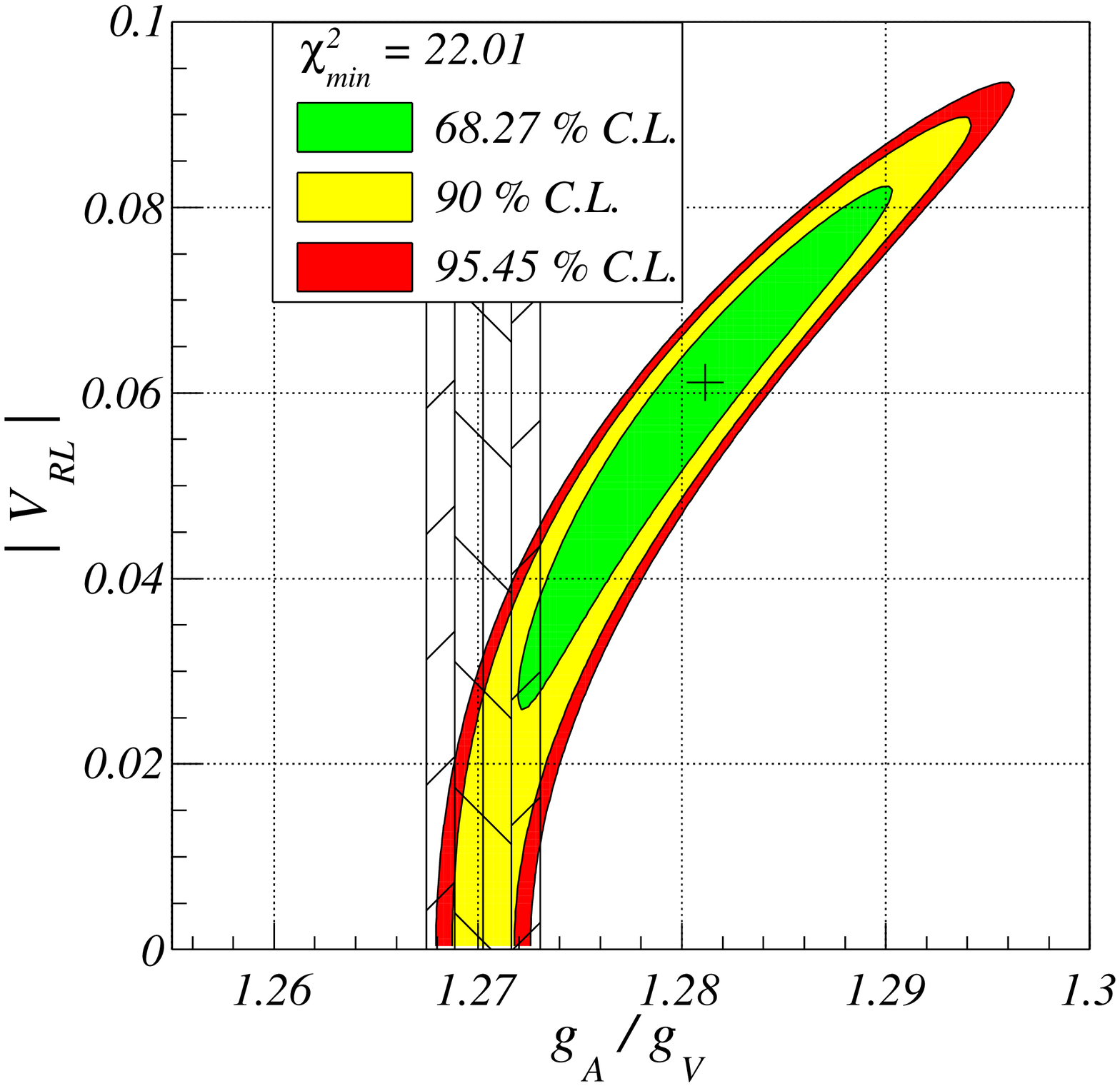}
\includegraphics[width=0.48\textwidth]{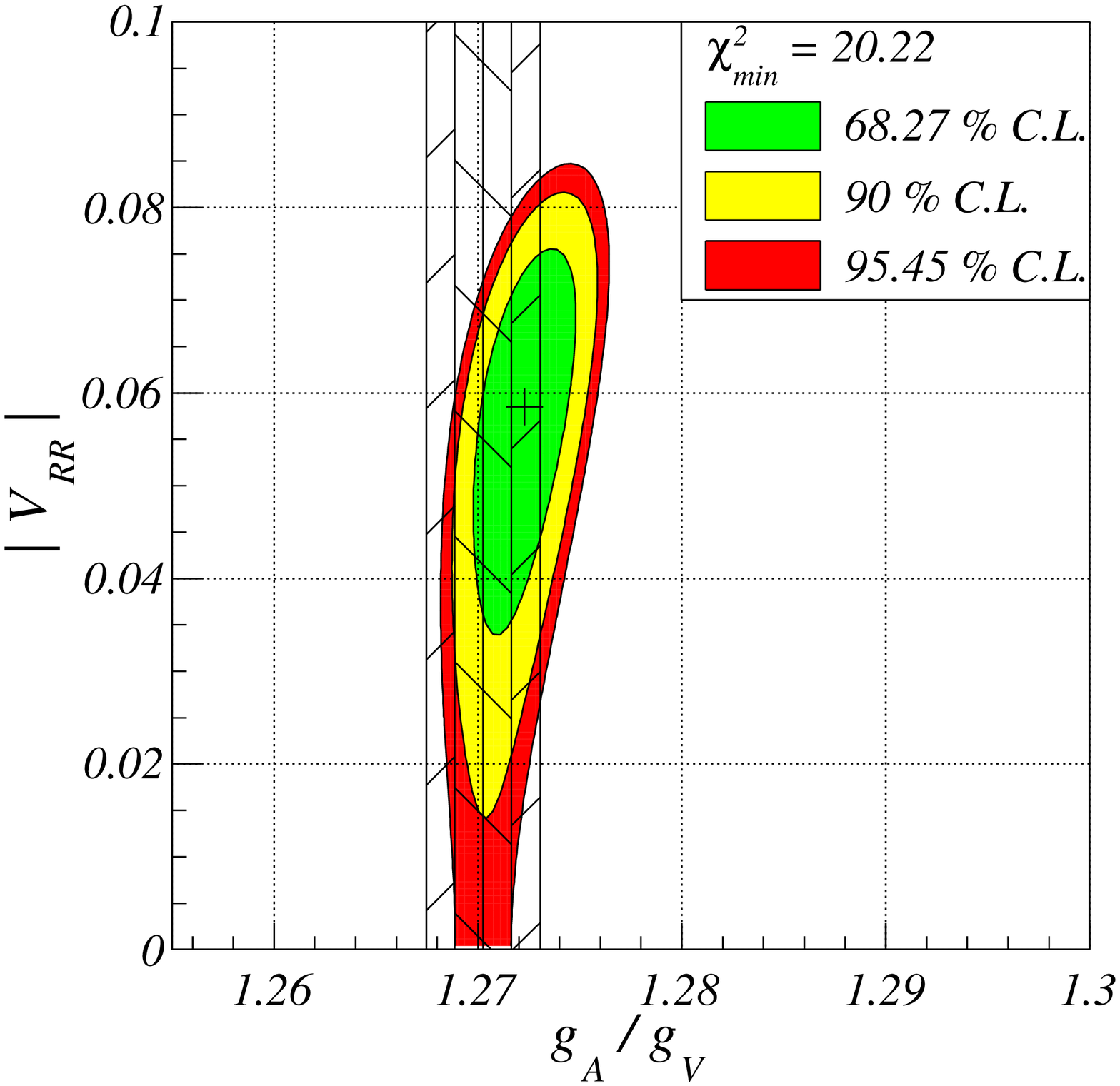}
\includegraphics[width=0.48\textwidth]{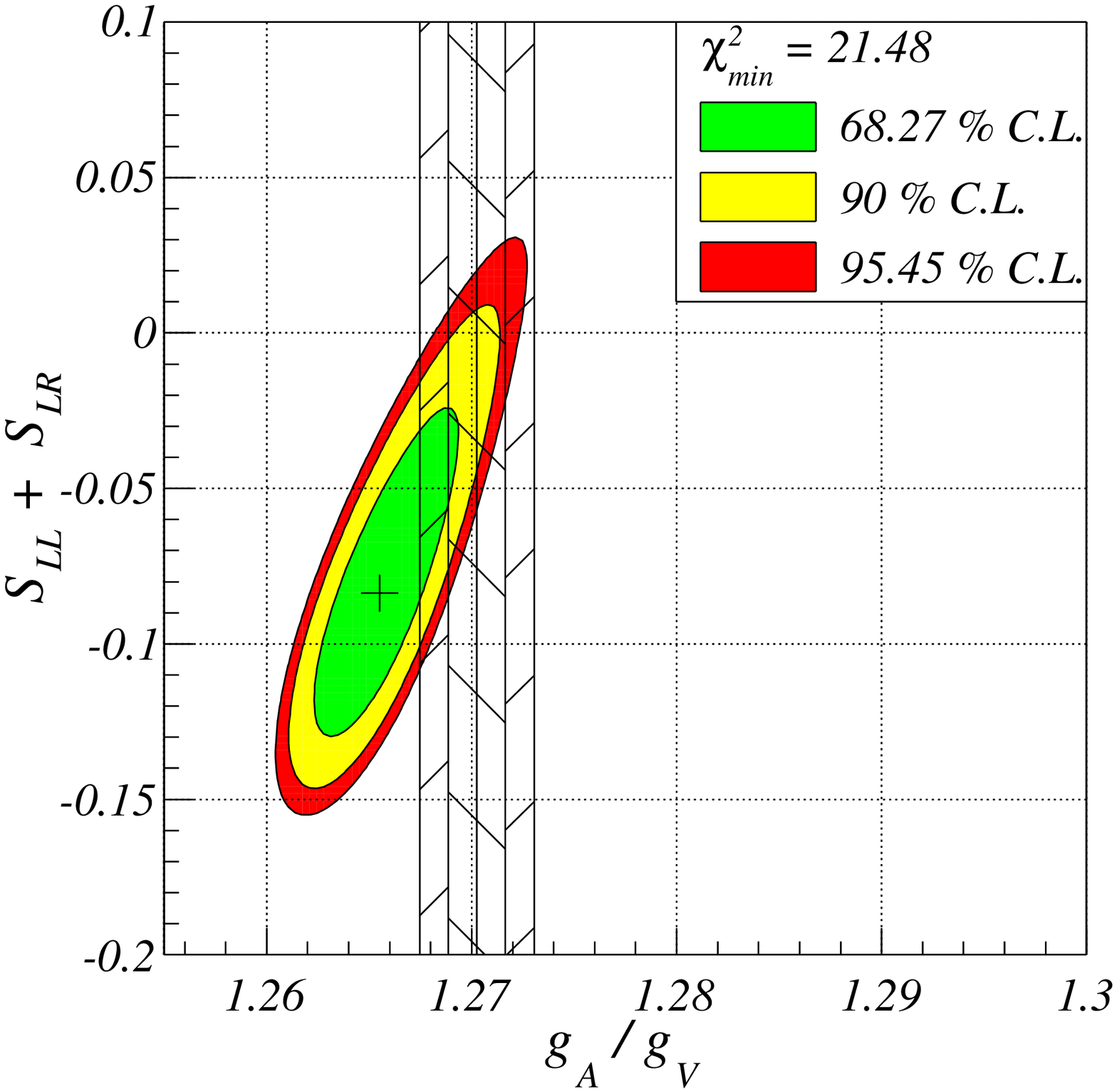}
\includegraphics[width=0.48\textwidth]{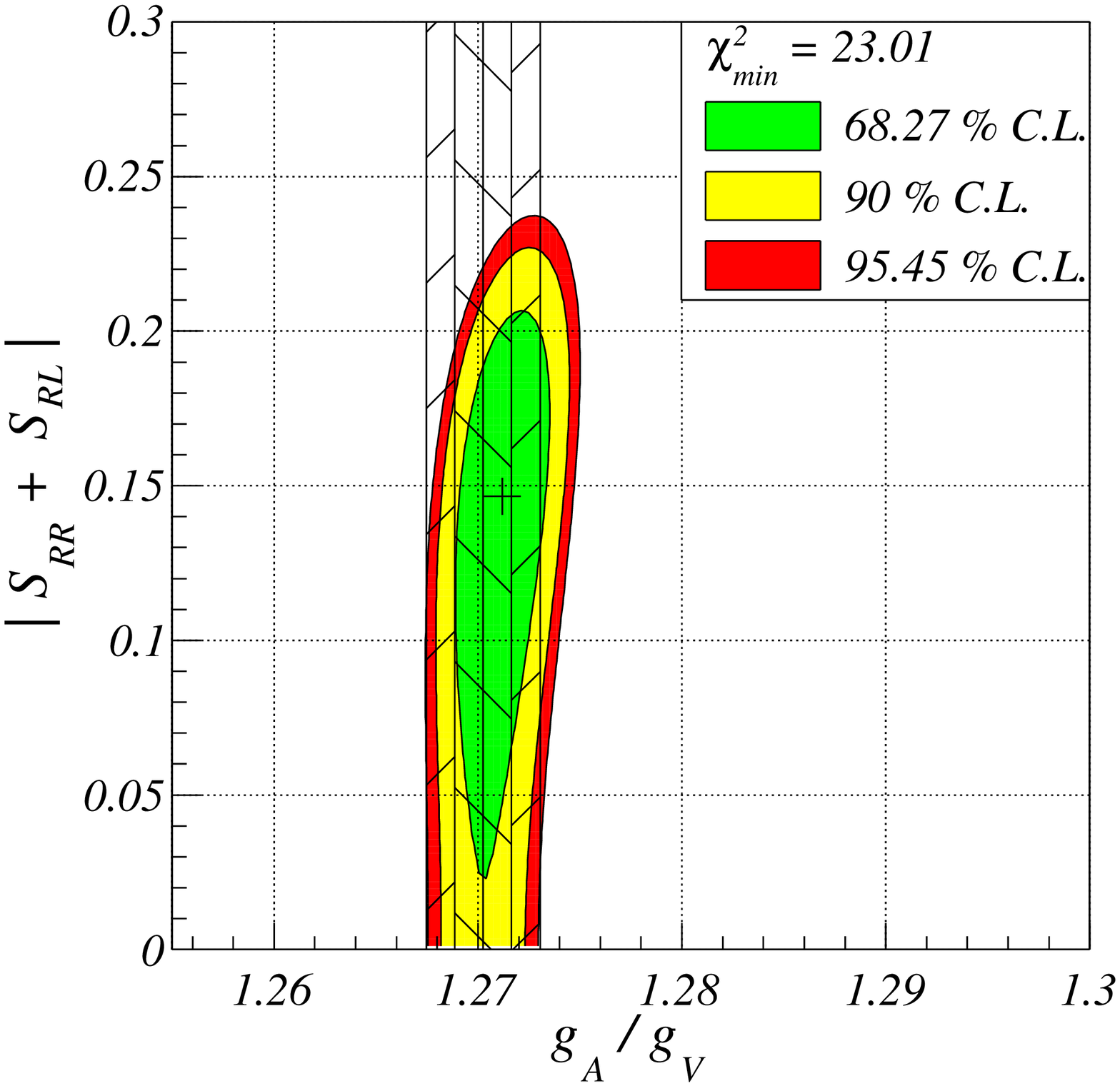}
\includegraphics[width=0.48\textwidth]{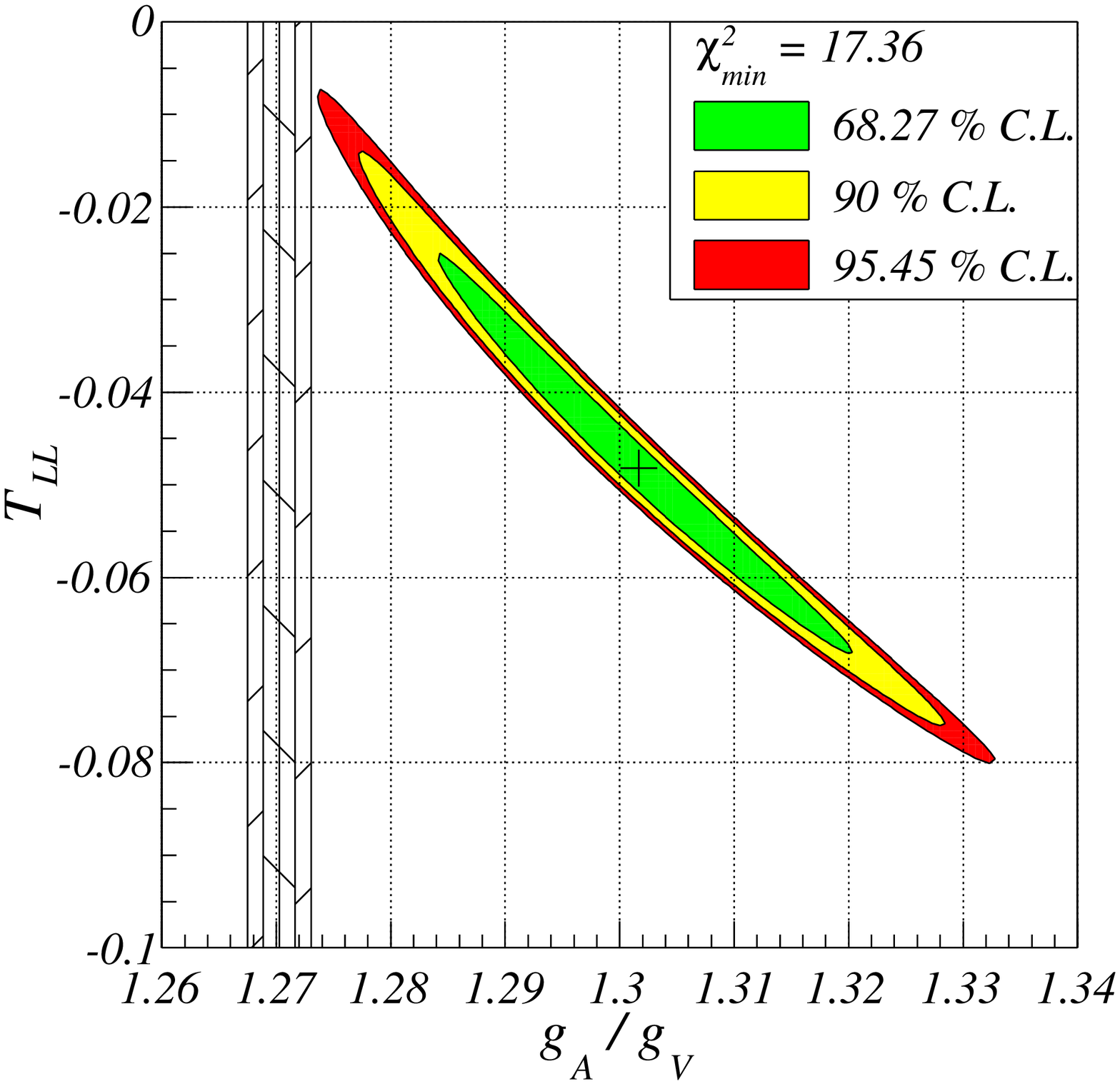}
\includegraphics[width=0.48\textwidth]{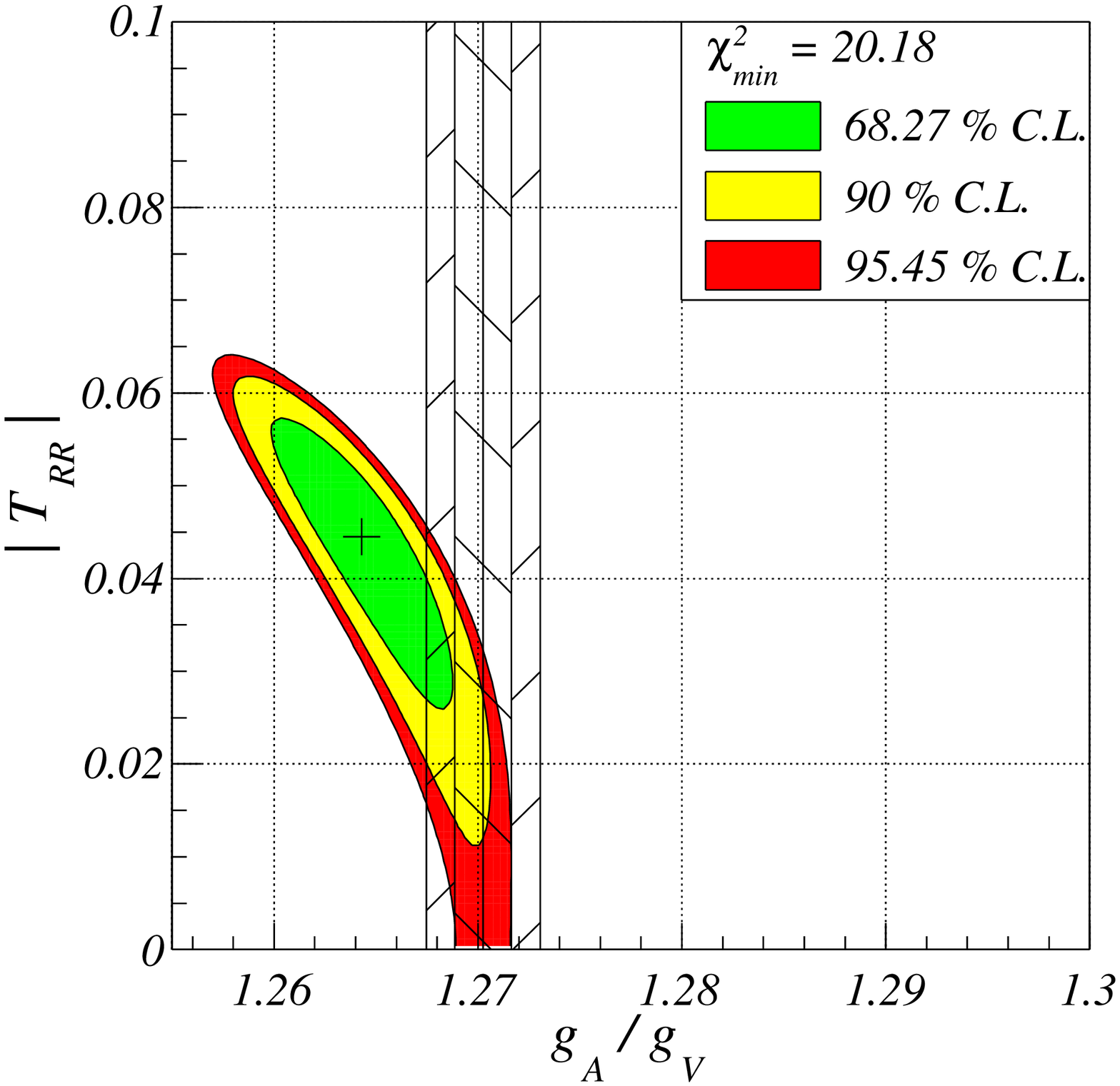}

\caption{Results of the two-parameter fits (the cross denote the position of $\chi^2_{min}$ in each case). The line --- marked areas correspond to the $\lambda\, =\, g_A/g_V$ intervals: (\ref{g_A_fit_68}) --- the narrow one and (\ref{g_A_fit_95}) --- the wider one. Note that $|\cdot|$ is the absolute value --- not the module of a complex number, as all parameters are real. \label{fits_2_param}}
\end{figure}

In the next step one of the parameters: $V_{Rk}$, $s_k$, $T_{kk}$ for $k = L,\,R$ is nonzero and fitted together with the ratio $g_A/g_V$. Among these cases only when the nonzero parameter is $s_L$ or $T_{LL}$ we have $b \neq 0$ and $b_\nu \neq 0$. The results of such two-parameter fits are presented in the Fig.~\ref{fits_2_param}. 
\\

In conclusion, Standard Model describes the neutron beta decay very well. The fits are minimally better if New Physics is included, especially if tensor terms are present. In some cases there is rather big dependence of the results on $g_{A}/g_{V}$ ratio. \\

This work has been supported by the Polish Ministry of Science and Higher Education under grant No. \mbox{N N202 064936}.

\end{document}